# MECHANICAL DESIGN OF CERAMIC BEAM TUBE BRAZE JOINTS FOR NOVA KICKER MAGNETS*

C.R. Ader[#], R.E. Reilly, and J.H. Wilson
Fermi National Accelerator Laboratory, Batavia, IL 60510, USA

## Abstract

The NOvA Experiment will construct a detector optimized for electron neutrino detection in the existing NuMI neutrino beam. The NuMI beam line is capable of operating at 400 kW of primary beam power and the upgrade will allow up to 700 kW. Ceramic beam tubes are utilized in numerous kicker magnets in different accelerator rings at Fermi National Accelerator Laboratory. Kovar flanges are brazed onto each beam tube end, since kovar and high alumina ceramic have similar expansion curves. The tube, kovar flange, end piece, and braze foil alloy brazing material are stacked in the furnace and then brazed. The most challenging aspect of fabricating kicker magnets in recent years have been making hermetic vacuum seals on the braze joints between the ceramic and flange. Numerous process variables can influence the robustness of conventional metal/ceramic brazing processes. The ceramic-filler metal interface is normally the weak layer when failure does not occur within the ceramic. Differences between active brazing filler metal and the moly-manganese process will be discussed along with the applicable results of these techniques used for Fermilab production kicker tubes.

## HISTORY

Ceramic beam tubes have always been used for kicker magnets at Fermilab. There are two reasons for using ceramic chambers in kickers: to avoid shielding of a fast changing external magnetic field by metallic chamber walls and to reduce heating due to eddy currents. The inner surfaces of the ceramic chambers are coated with a conductive coating which has a graphite base to reduce the beam coupling impedance and provide passage for beam image current. The tubes must meet requirements including length and straightness in order to fit into the magnet, and also have sufficient strength to withstand atmospheric and mechanical forces encountered during assembly and operation.

The ceramic beam tubes have been found to the largest technical risk for the kicker magnets in all the recent accelerator projects at Fermilab. The cost per brazed tube has been $33,625 [1] which exceeds the cost of the rest of the entire kicker magnet. Although the brazed ceramic beam tube is a fairly straightforward component of the kicker magnet, it has typically been on the critical path of the fabrication since it is so susceptible to leaks and has had such a high failure rate.

For example, recently produced rectangular tubes for the Fermilab Booster Ring have had less than a 20% vacuum-leak tightness rate using the active brazing process. Argonne National Laboratory has had similar problems and they eventually had to epoxy the flanges to the kicker tubes because of the difficulties encountered [2].

## DESIGN

An elliptical shaped tube is used in order to give the proper beam aperture and also adequate strength for the internal vacuum pressure (Figure 1). The wall thickness used is approximately 4.8 mm, although for other projects the wall thickness has been thinner.

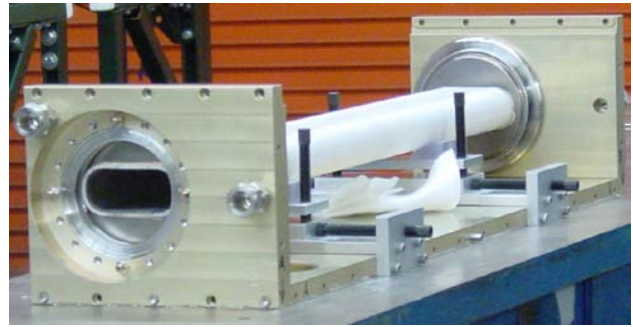

Figure 1: Brazed ceramic beam pipe during the welding process to end bellows.

Ceramic tubes produced for other Fermilab projects, such as the Booster Accelerator, have included other shapes, including rectangular. Depending on the beam aperture and beam-line requirements which dictate the magnet size, the inner and outer diameter of the tube sizes are optimized.

## FABRICATION OF COMPONENTS

The kovar flanges are fabricated in-house and are made of 0.381 mm thick Kovar, ASTM-F-15, MIL-1-23011 CL 1, cold rolled, which are first roughly cut to size and then hydroformed using a die.

In order to make a tube that is 2.0 m long, a mandrel of 3.1 m has to be used in order to compensate for the fact that the ceramic shrinks about 22% and there is about 203 mm of tube waste on each end due to manufacturing techniques. Ceramic tubes are fabricated using a slip cast method using a straight tapered mandrel, with a taper of approximately 0.013 degrees from straight all around in the 3.3 m length. McDanel Advanced Ceramic Technologies have fabricated the tubes which are 99.8% alumina.

*Work supported by Fermi Research Alliance, LLC, under Contract No. DE-AC02- 07CH11359 with the U.S. Department of Energy.
[#]cader@fnal.gov

## CERAMIC-TO-KOVAR JOINT DESIGN

Kovar is used to braze to the ceramic because the coefficients of expansion of the ceramic and kovar are closely matched. Nickel and nickel alloys can also be used because they are very ductile and are able to absorb the stresses. Kovar is an alloy of 29% Ni, 17% Co, 0.30% Mn, with iron making up the balance.

Before brazing, lapping has been used on the tube ends to reduce the number of near-surface micro-cracks. This surface damage can increase the probability that residual stresses will produce catastrophic cracks during the cooling of ceramic-to-metal joints, and can reduce the load-carrying capability of both ceramic and ceramic-to-metal joints. Lapping is a process that uses a fine, loose adhesive in a fluid suspension to produce very fine surface texture and a high degree of flatness. The ceramic is more polished when the ends are lapped with less grain pullout due to the uniformity and higher density [3].

The kovar disc on the tube is typically dished in order to take the radial expansion and contraction forces of the flange on the tube during the brazing process. This design works well with metallization. However, it was found that a flat kovar flange configuration needs to be used with the active brazing process because it requires a flatter surface in order to make better surface contact (Figure 2).

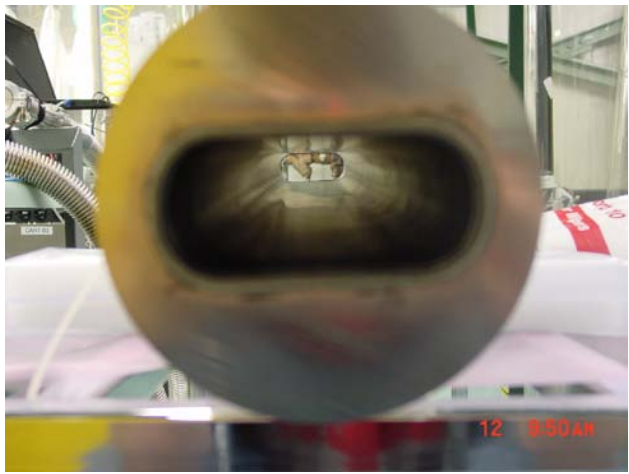

Figure 2: End view of a flat kovar flange brazed on a ceramic tube.

The flange is made with an inner diameter about 6.35 mm too small so that there will be about 3.2 mm lip all around the inner diameter of the tube. This provides some tolerance if the tube moves during the brazing process and also more surface area for the braze material. This inner lip is ground down after brazing in order to match the inner diameter of the ceramic tube.

### Active Brazing

The tube, flange, end piece and a braze foil made of titanium and inusel are stacked in the furnace with about 2.7 kilogram of weights to hold them together when they are in the furnace. The tube is brazed to about 850° C. Typically one end is brazed at a time in a horizontal oven.

During the brazing process and subsequent cool-down to room temperature, residual stresses are generated due to the differential thermal contractions of the alumina, silver braze and kovar flanges [4]. As the joint is cooled from the brazing temperature to ambient temperature, the interface microstructure changes and residual stresses develop. Sometimes these braze joints subsequently will crack and will not be leak tight (Figure 3).

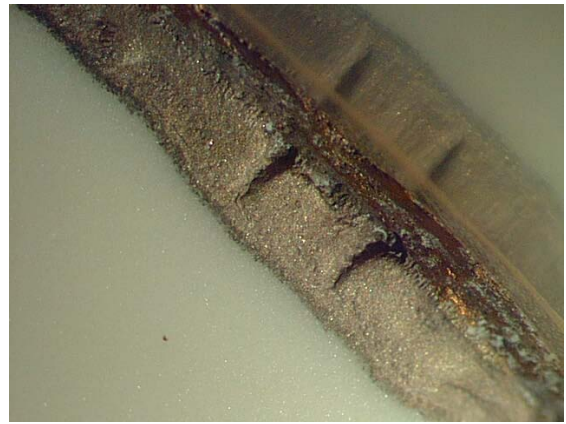

Figure 3: Active braze joint between the ceramic tube and kovar flange which has cracked.

Active brazing is used on parts which are too large to fit in the ovens for metallization. Metallization brazing actually needs two furnaces: one for metalizing the ceramic and another for brazing the kovar-to-ceramic.

### Moly-Manganese Brazing

One of the most widely used methods of metallization is the moly-manganese process which is used extensively in the electronic and electrical industries to produce ceramic-to-metal seals for ultra-high vacuum equipment such as isolators and power semiconductor housings.

The major difference between active brazing and metallization is the adhesion strength of the active braze is between 30 and 49 MPa. The strength of the metalized parts are between 49 and 98 MPa.

Factors which would increase the success of brazing would be to increase the tube wall thickness, larger grain sizes in the ceramic, and a more uniform grain structure in the ceramic. Additionally, a denser grain structure caused by higher firing temperatures would help. Recent tubes fabricated by McDanel Advanced Ceramics have a wall thickness that is almost 50% smaller than in the past. This is a factor why there have been difficulties in brazing [5].

Active brazing has been used because no vendors have been found which can metallize long tubes. However, Kyocera in Japan can metallize long tubes, up to 1.5 m. Ceramic is metallized with 80% molybdenum and 20% manganese powder of one to five micron particles in a binder which is burned off during firing at 1500° C in a wet hydrogen furnace. The metal powder sinters, penetrates the ceramic surface, and bonds with the ceramic, forming a coating 0.025 to 0.038 mm thick.

A second coating is needed to get the solder to wet the metallized area during brazing. The metallized area is then nickel plated it a thickness of 0.003 mm, then re-fired in a reducing atmosphere of 1000° C. The nickel coating enhances wetting and facilitates leak-tight joints. The piece is then brazed using a conventional braze, such 72% silver and 28% copper eutectic alloy and brazed at about 780° C.

## VENDOR CONSIDERATIONS

A search has been conducted worldwide for vendors using other National Labs as contacts such as CERN, DESY, Argonne National Laboratories, Jefferson Laboratories, and others. A balance between brazing methods and other quality issues has had to be taken. Fermilab's preference is to do metallization because of the 100% leak-tight success rate. The success with active brazing is less than 50%.

Other Labs such as Los Alamos have fabricated tubes with the tube made in two pieces, a top and bottom [6]. The ceramic tubes are then brazed with a glass frit mixture which is basically crushed glass. Fermilab investigated a similar idea using brazing with glass frit which was not successful.

## OTHER CONSIDERATIONS

One problem that has been encountered with the metallization process has been discoloration of two tubes afterwards. This may be due to recrystallization during the furnace brazing process [7]. Some of the discoloration is due to recrystallization of glass phase in the ceramic. It leaves the appearance of multiple small spots on the surface. This is not so much contamination as it is a different microstructure. However, in the past, different phases have been seen that have been more or less susceptible to chemical attack. The standard firing processes may have caused the glass crystalline formation. Then subsequent chemical processing through nickel plating and cleaning may have attacked the susceptible phase, possibly causing entrapment of plating/cleaning chemicals. The firing processes helps to reveal any chemicals entrapped.

Another problem has been the cracking of the kovar flange. This typically has occurred during active brazing and Sandia Laboratories observed fractures in wire that had abnormally large grain size. They found that if the kovar sheet had large grains to start with or the annealing or the brazing temperatures grew the grains large, this may cause the problem [8].

## SUMMARY

Seven shorter Recycler Gap Clearing Kicker magnets were installed last year (Figure 4). A purchase requisition for the brazing of twelve 1219 mm long tubes has been written and approved. Six tubes will be brazed by one vendor and the other six by another. The magnet design has been completed and the plan is that seven Recycler Extraction, Main Injection, and Beam Abort kickers will be ready for installation in early 2012.

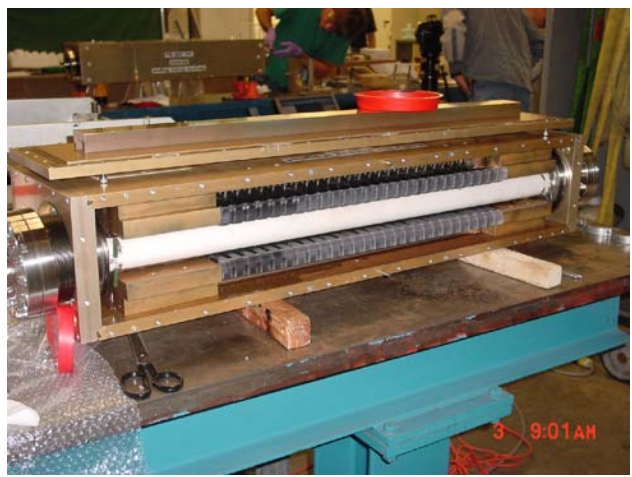

Figure 4: Ceramic tube installed in kicker magnet with one of the side plates removed.

## ACKNOWLEDGEMENTS

We wish to thank Dirk Hurd, Eric Pirtle, and Rod Stewart (CAD Support); B. Tennis and F. Juarez (Technician Support); and also B. Jensen, A. Makarov, and O. Kiemschies (Technical Division). Special thanks to C.C. Jensen.